\documentclass[%
 aip,
 jmp,%
 amsmath,amssymb,
 preprint,%
]{revtex4-1}

\usepackage{stmaryrd}
\usepackage{graphicx}
\usepackage{bm}
\usepackage{textcomp}
\usepackage{subfigure}
\usepackage{psfrag}
\usepackage{multirow,amssymb,amsbsy,amsmath}
\usepackage{ctable}
\usepackage{colortbl}
\usepackage{CJK}
\usepackage{amsmath}

\begin{document}

\title{Probing and modelling the localized self-mixing in a GaN/AlGaN field-effect terahertz detector}

\author{J. D. Sun}
\affiliation{Key Laboratory of Nanodevices and Applications, Suzhou Institute of Nano-tech and Nano-bionics, Chinese Academy of Sciences, 398 Ruoshui Road, Suzhou, Jiangsu 215123, P.~R.~China}
\affiliation{Institute of Physics, Chinese Academy of Sciences, Beijing 100190, P.~R.~China}
\affiliation{Graduate University of Chinese Academy of Sciences, Beijing 100049, P.~R.~China}

\author{H. Qin}\email{hqin2007@sinano.ac.cn}
\affiliation{Key Laboratory of Nanodevices and Applications, Suzhou Institute of Nano-tech and Nano-bionics, Chinese Academy of Sciences, 398 Ruoshui Road, Suzhou, Jiangsu 215123, P.~R.~China}

\author{R. A. Lewis}
\affiliation{Institute for Superconducting and Electronic Materials, University of Wollongong, Wollongong, New South Wales 2522, Australia}

\author{Y. F. Sun}
\affiliation{Key Laboratory of Nanodevices and Applications, Suzhou Institute of Nano-tech and Nano-bionics, Chinese Academy of Sciences, 398 Ruoshui Road, Suzhou, Jiangsu 215123, P.~R.~China}
\affiliation{i-Lab, Suzhou Institute of Nano-tech and Nano-bionics, Chinese Academy of Sciences, 398 Ruoshui Road,
Suzhou, Jiangsu 215123, P.~R.~China}

\author{X. Y. Zhang}
\affiliation{Key Laboratory of Nanodevices and Applications, Suzhou Institute of Nano-tech and Nano-bionics, Chinese Academy of Sciences, 398 Ruoshui Road, Suzhou, Jiangsu 215123, P.~R.~China}

\author{Y. Cai}
\affiliation{Key Laboratory of Nanodevices and Applications, Suzhou Institute of Nano-tech and Nano-bionics, Chinese Academy of Sciences, 398 Ruoshui Road, Suzhou, Jiangsu 215123, P.~R.~China}

\author{D. M. Wu}
\affiliation{Key Laboratory of Nanodevices and Applications, Suzhou Institute of Nano-tech and Nano-bionics, Chinese Academy of Sciences, 398 Ruoshui Road, Suzhou, Jiangsu 215123, P.~R.~China}
\affiliation{i-Lab, Suzhou Institute of Nano-tech and Nano-bionics, Chinese Academy of Sciences, 398 Ruoshui Road,
Suzhou, Jiangsu 215123, P.~R.~China}

\author{B. S. Zhang}
\affiliation{Key Laboratory of Nanodevices and Applications, Suzhou Institute of Nano-tech and Nano-bionics, Chinese Academy of Sciences, 398 Ruoshui Road, Suzhou, Jiangsu 215123, P.~R.~China}

\date{\today}

\begin{abstract}
In a GaN/AlGaN field-effect terahertz detector,
the directional photocurrent is mapped in the two-dimensional space of the gate voltage and the drain/source bias.
It is found that not only the magnitude, but also the polarity, of the photocurrent can be tuned.
A quasistatic self-mixing model taking into account the \textit{localized} terahertz field
provides a quantitative description of the detector characteristics.
Strongly localized self-mixing is confirmed.
It is therefore important to engineer the spatial distribution of the terahertz field and
its coupling to the field-effect channel on the sub-micron scale.
\end{abstract}

\keywords{terahertz detector, plasma wave, terahertz antenna, high electron mobility transistor}
\maketitle

Field-effect transistors (FET) based on high-mobility two-dimensional electron gas (2DEG) are promising candidates
for room-temperature high sensitivity terahertz detectors.~\cite{dyakonov-ieee96,knap-jap2002,knap-apl2002,veskler-prb2006,lisauskas-JAP2009,glaab-apl2010}
The related Dyakonov-Shur's non-quasistatic theory for
the plasmon-induced photo voltage/current has been studied extensively in the resonant and non-resonant regimes.~\cite{dyakonov-ieee96}
In the former regime, the channel length $L$ is shorter than the
plasma-wave decay length ($\omega\tau \gg 1$, $s\tau \gg L$),
where $\omega$ is the plasma frequency, $\tau$ is the momentum relaxation time of travelling electrons,
$s$ is the plasma wave velocity.
A resonant detection requires a short electron channel and
a special drain/source boundary condition to enhance the $Q$ factor of the plasma-wave cavity.
The experimentally observed resonant detection is usually realized
in nanometer-sized FETs at cryogenic temperatures.~\cite{knap-apl2002,knap-jap2002,veskler-prb2006}
In contrast, a non-resonant detection ($\omega\tau \lesssim 1$, $s\tau \ll L$)
relies on the rectification/self-mixing of damped plasma waves.
As has been pointed out in Ref.~\onlinecite{dyakonov-ieee96,lisauskas-JAP2009},
a self-mixing detector can deliver a high responsivity in a broad frequency range far above
the transistor's cut-off frequency, even in a long-channel FET at room temperature.
In the self-mixing model, a terahertz excitation is usually assumed
at the edges of the gate with the source being grounded.
Ref.~\onlinecite{lisauskas-JAP2009} studied the
spatio-temporal carrier dynamics in the electron channel.
The simulation showed that the plasma wave decays rapidly from the excitation point toward the
other end of the channel with a decay length less than 50~nm for silicon at room temperature.
It is within such a short distance where the terahertz photovoltage is built up.
The effect of spatial distribution of electron density
tuned from the linear regime to the saturation regime by the drain/source bias ($V_{DS}$) and the gate voltage ($V_G$)
has been studied both theoretically and experimentally.~\cite{veskler-prb2006,lisauskas-JAP2009,elkhatib-apl2011}
However, a complete map of the terahertz response in the 2D space of $V_{DS}$ and $V_G$ is not yet available.
It is of great utility to further reveal the spatial distribution of the self-mixing,
so that a more accurate device model may be constructed for detector optimization.
In this letter, we report a full characterization of a self-mixer
by mapping out the photocurrent in a 2D space of $V_{DS}$ and $V_G$,
covering the linear (LR) and saturated regimes (SR) and the transition regime (TR) in between.
A technique to reveal the inhomogeneous distribution of terahertz field and
the resulting self-mixing is introduced.
A quasistatic self-mixing model is constructed and examined,
showing a good quantitative fit to the experiment results.


To focus on the device physics, we present the experiment data obtained
in a GaN/AlGaN field-effect terahertz detector similar to that reported in our previous work.~\cite{jdsun-apl2012}
We have to mention that similar results were also observed in other detectors
with different antennas or a different channel length.~\cite{note_1}
A scanning-electron micrograph of the detector and the measurement setup are shown in Fig.~\ref{fig_1}(a)
The key features of the detector are (1) a 2-$\mathrm{\mu m}$ field-effect gate is connected to
an antenna block (\textit{g}-antenna), (2) two isolated antenna blocks (\textit{i}-antenna)
sit by the gate with an equal gap of $1.5~\mathrm{\mu m}$,
(3) the drain and source ohmic contacts are set about $64~\mathrm{\mu m}$ away from the gate.
At 77~K, the electron mobility and the density are $\mu = 1.58\times 10^4~\mathrm{cm^2/Vs}$
and $n_s=1.06\times 10^{13}~\mathrm{cm^{-2}}$, respectively.
The width of the electron channel (mesa) is $W=4~\mathrm{\mu m}$.
More details of the detector have been presented in Ref.~\onlinecite{jdsun-apl2012}.
The terahertz wave, generated from a backward-wave oscillator (BWO: $f=\omega/2\pi=903$~GHz),
is collimated and focused onto the detector cooled at 77~K.
The polarization of the terahertz wave is parallel to the i-antennas.
As shown in Fig.~\ref{fig_1}(a),
we used a standard lock-in technique, in which the terahertz wave is
modulated by a chopper at frequency $f_M=317$~Hz.
The directional photocurrent is calculated from the magnitude and the phase of
the AC voltage across a sampling resistor (10~$\Omega$).
The DC voltage source ($V_S$ or $V_D$) and the sampling resistor are connected to the drain electrode
and the source electrode, respectively, or vice versa.
The order of the lock-in amplifier's (LIA-S or LIA-D) differential inputs (A and B)
is kept consistent with the direction of a positive photocurrent.
The detector is first characterized without terahertz irradiation.
As shown in Fig.~\ref{fig_1}(b), the current-voltage curves exhibit
the standard characteristics of a FET.
According to the gradual-channel approximation (GCA),
the channel current in the linear regime (LR: $V_{DS}<V_G-V_T$) is written as
\begin{equation}
i_x = e\mu Wn_x \frac{\mathrm{d}V_x}{\mathrm{d}x}, \label{eq_1}
\end{equation}
where, the channel potential $V_x$ varies from 0 to $V_{DS}$ from $x=0$ to $x=L$,
and $V_T$ is the threshold gate voltage.
The local electron density is controlled by the effective local gate voltage: $n_x=C_GV_{GE}/e=C_G(V_G-V_T-V_x)/e$,
where $C_G$ is the effective gate-channel capacitance.
By measuring the differential conductance
($G_0=\mathrm{d}I_{DS}/\mathrm{d}V_{DS} \approx \mu WC_G(V_G-V_T)/L$) at $V_{DS}=0$~V
as shown in Fig.~\ref{fig_1}(c),
the threshold voltage and the gate capacitance are estimated to be
$V_T \approx -4.34$~V and $C_G \approx 0.4~\mathrm{\mu F/cm^2}$, respectively.
In the GCA model, the drain-source current and the local channel potential can be written as
\begin{widetext}
\begin{equation*}
I_{DS}=
\begin{cases}
\mu C_GW\big[ 2(V_G-V_T)V_{DS}-V_{DS}^2 \big]/2L & \text{if } V_{DS} \leq V_G-V_T, \\
\mu C_GW\big[ (V_G-V_T)^2+\lambda(V_{DS}-V_{G}+V_T) \big]/2L & \text{if } V_{DS}>V_G-V_T,
\end{cases}
\end{equation*}
\begin{equation}
V_x=
\begin{cases}
(V_G-V_T)[1-(1-x/L_{LR})^{1/2}], x=[0, L] & \text{if } V_{DS} \leq V_G-V_T, \\
(V_G-V_T)[1-(1-x/L_{SR})^{1/2}], x=[0, L_{SR}] & \text{if } V_{DS}>V_G-V_T,
\end{cases}~\label{eq_2}
\end{equation}
\end{widetext}
where, parameter $\lambda \approx 0.08$~V describes the degree of the effective channel-length modulation
($L \rightarrow L_{SR}=L/[1+\lambda(V_{DS}-V_G+V_T)/(V_G-V_T)^2]$) in regime SR,
and $L_{LR}=L(V_G-V_T)^2/[2(V_G-V_T)V_{DS}-V_{DS}^2]$ in regime LR.
From Eq.~\ref{eq_2} and the experimental data shown in Fig.~\ref{fig_1}(c),
the charge density and its derivative $\mathrm{d}n/\mathrm{d}V_{GE}$
at different locations can be numerically calculated.


Upon terahertz irradiation with a frequency of $\omega$ and an energy flux of $P_0$,
both a horizontal ($\dot\xi_xE_0$) and a perpendicular ($\dot\xi_zE_0$) terahertz field in the channel are induced.
Here, the free-space field strength $E_0$ is determined from
$P_0$ and the free-space impedance $Z_0$ by $P_0=E_0^2/2Z_0$.
Parameter $\dot\xi_x$ and $\dot\xi_z$ stand for
the field enhancement factors for the horizontal and perpendicular field, respectively.
As shown in Fig.~\ref{fig_2}(a,b) from a FDTD simulation,
the horizontal field is concentrated in the gaps between the gate and the i-antennas,
while the perpendicular field is mainly distributed under the gate.
Both fields are stronger at the source side of the gate than at the drain side.
The horizontal field vanishes and changes its phase by $180^\circ$ at $x_c \approx +0.3~\mathrm{\mu m}$,
while the perpendicular field keeps its phase constant along the channel (Fig.~\ref{fig_2}(c,d)).
Since the terahertz fields are highly localized within
a distance of 200~nm to the edges of the gated area and the phases evolve in a continuous manner,
the effective electron channel can be modeled as a chain of nanometer-sized FETs excited by a spatially decaying terahertz field.
At 77~K, the momentum relaxation time is about $\tau \approx 2$~ps,
which corresponds to a mean-free-path of $\lesssim 1~\mathrm{\mu m}$.
In spite of the fact that $\omega\tau \approx 10$,
in each nanometer-sized transistor, the electron transit time ($\tau_D \ll 200~\mathrm{nm}/v_D \approx 2$~ps)
and the plasma-wave transit time ($\tau_p \ll 200~\mathrm{nm}/s \approx 0.2$~ps)
become smaller than the period of the oscillating field, i.e., $\omega\tau_{D/p} \ll 10$.
In the following,
we will present the map of terahertz response in the 2D space of $V_{DS}$ and $V_G$
and verify a quasistatic self-mixing model taking into account the non-uniform terahertz excitation.
Under the quasistatic assumption,
the local effective channel potential and the effective gate voltage can be written as
$V_x \rightarrow V_x + \xi_x E_0\cos(\omega t+\phi_x)$ and
$V_G \rightarrow V_G + \xi_z E_0\cos(\omega t+\phi_z)$ with
$\mathrm{d}\xi_x/\mathrm{d}x = \dot\xi_x$ and $\mathrm{d}\xi_z/\mathrm{d}z = \dot\xi_z$, respectively.
By substituting the above quantities into Eq.~\ref{eq_1} and
integrating along the channel, we obtain the DC terahertz photocurrent
\begin{eqnarray*}
i_T &=& i_{xz}+i_{xx} \\
&=& \frac{e\mu W}{2L}Z_0P_0 \int_0^L \frac{\mathrm{d}n}{\mathrm{d}V_{GE}} \big[ \dot \xi_x\xi_z\cos\phi - \dot \xi_x\xi_x \big] \mathrm{d}x,
\end{eqnarray*}
where $\phi=\phi_x$ since $\phi_z=0$.
Current $i_{xz}\propto \dot{\xi_x}\xi_z\cos\phi$ is induced by both horizontal and perpendicular fields, while
$i_{xx}\propto \dot{\xi_x}\xi_x$ is only from the horizontal field.
For numerical simulation, we approximate the term $i_{xz}$ as
\begin{equation}
i_{xz} = \frac{e\mu W}{2L}Z_0P_0 \bar{z}\int_0^L \frac{\mathrm{d}n}{\mathrm{d}V_{GE}}\dot\xi_x\dot\xi_z\cos\phi \mathrm{d}x,
\end{equation}
where, $\bar{z}=\xi_z/\dot{\xi}$ is the effective distance between the channel and the gate.
Due to the phase change at $x_c$, mixing term $i_{xz}$ changes
from a positive value at the left side to a negative value at the right side of $x_c$.
In the case of the antenna design shown in Fig.~\ref{fig_1},
term $i_{xz}$ is the major contribution to the induced photocurrent.~\cite{note_2}


In Fig.~\ref{fig_3}(a),
the photocurrent as a function of $V_{DS}$ and $V_G$ is mapped in a color-scale plot.
For comparison, a simulated map based on the above quasistatic model is shown in Fig.~\ref{fig_3}(b).
In the simulation, we used two sets of input data.
One of them includes the parameters such as the electron mobility, the gate geometry,
the threshold voltage and the gate-controlled electron density obtained from Fig.~\ref{fig_1}(c).
The other set of input data includes the spatial distribution
of the terahertz fields obtained from the FDTD simulations, as shown in Fig.~\ref{fig_2}.
The simulation reproduces the main features of the polarity change and the magnitude variation.
In regime PR, the whole channel is pinched off and the terahertz mixing is disabled.
In the linear regime, there is only a very weak photocurrent.
A strong photocurrent is produced in regime TR and SR, and strongly depends on the applied bias and gate voltage.
When the bias is applied at the drain,
a negative photocurrent is flipped into a stronger positive photocurrent
when the gate voltage is swept from regime TR into regime PR.
In contrast, when the bias is applied at the source
and the gate voltage is swept, a positive current is flipped into a weaker negative current.
When the bias is swept and switched to either the drain or the source with a constant gate voltage at -4.1~V,
the photocurrent differs in polarity and its magnitude becomes fairly independent of the bias voltage.


The spatial distributions of the terahertz fields ($\dot\xi_x\dot\xi_z\cos\phi$) and
the derivative of the charge density ($\mathrm{d}n/\mathrm{d}V_{GE}$)
at $V_G=-4.1$~V and $V_D=2.0$~V or $V_S=2.0$~V are plotted in Fig.~\ref{fig_4}(a).
Due to the diminishing of charge modulation at the side where a positive bias is applied, the local self-mixing is suppressed.
Hence, a polarity change is observed between the cases of $V_D=2$~V and $V_S=2$~V.
To further illustrate the terahertz responses in different regimes,
we mark three lines ($V_S=0$~V, $V_D=V_S=0$~V, $V_D=2$~V) on the maps shown in Fig.~\ref{fig_3}.
Along the line at $V_D=2$~V, we select five conditions labeled as
$\alpha$, $\beta$, $\gamma$, $\delta$, and $\chi$,
corresponding to $V_G=-1.0, -2.0, -3.0, -3.5$, and -4.1~V, respectively.
The location-dependent charge modulation is plotted in Fig.~\ref{fig_4}(b).
By decreasing the gate voltage from $\alpha$ to $\chi$,
the charge modulation becomes stronger and the location of the maximum
shifts from the drain side to the source side.
This evolution gives rise to the observed transition from a weak negative current
to a strong positive current along the line at $V_D=2.0$~V.
In Fig.~\ref{fig_4}(c-e),
the experimental gate-controlled photocurrent is compared to the simulation.
At zero bias, the photocurrent resembles what we observed in our previous studies.~\cite{yfsun-apl2011, jdsun-apl2012}
At $V_D=2.0$~V, the simulation fits well to the experimental data.
As shown in Fig.~\ref{fig_4}(e), when the bias is applied at the source side,
a transition from positive current at $V_G \approx -2$~V to a weaker negative current at $V_G \approx -4.1$~V is reproduced.
There is, however, a strong suppression of photocurrent
from $V_G=-3.5$~V to $V_G=-2.5$~V (also marked as vertical double arrows in Fig.~\ref{fig_3}),
while according to the quasistatic model a strong positive current should occur.
The absence of photocurrent in region SR is also observed in our detectors described in
Ref.~\onlinecite{yfsun-apl2011} and other detectors with symmetric antennas.~\cite{note_2}
The common feature is that the self-mixing in regime SR is suppressed
as long as the electron density at the same side of the strong terahertz field is set to be mostly depleted.
The deviation between the experiment and the model indicates that
there is a different transport mechanism.
The most likely candidate is the plasma-wave induced photocurrent.
To reveal the physics behind,
we will conduct systematic experiments at lower temperatures on different antenna designs to
tune both the amplitude and the phase of the terahertz field.
Accordingly, a non-quasistatic model taking into account the localized terahertz excitation will be examined.

In conclusion, the terahertz response in a GaN/AlGaN field-effect detector was mapped.
We directly probed the local self-mixing by selectively depleting the 2DEG near the edges of the channel.
A quasistatic model was constructed and well describes the main features of the terahertz response.
The finding confirms that non-uniform terahertz fields and hence self-mixing are induced in the channel.
The model provides valuable information as to how to construct a high responsivity terahertz field-effect detector.
The deviation between the experiment and the model suggests that a full model taking into account the
dynamic transport and the spatial distribution of terahertz excitation is desirable.

The authors gratefully acknowledge the support from the National Basic Research Program of China (G2009CB929303),
the Knowledge Innovation Program of the Chinese Academy of Sciences (Y0BAQ31001),
and the National Natural Science Foundation of China (60871077).
R.A.L. acknowledges support from the Chinese Academy of
Sciences Visiting Professorship Program
of Senior International Scientists (2010T2J07).


\newpage
\begin{figure}[!p]
\caption{
(a) Measurement circuit diagram.
The inset is an artificially-colored scanning-electron micrograph of the detector.
(b) The $\textit{I-V}$ characteristics of the detector.
(c) The differential conductance at $V_{DS}=0$~V as a function of $V_G$.
(d) A color-scale plot of the differential conductance as a function of $V_G$ and
drain/source bias ($V_D$ or $V_S$) which is applied either at the drain or at the source.
The dashed lines separate the map into the linear regime (LR), the transition regime (TR),
the saturation regime (SR), and the pinch-off regime (PR).
}\label{fig_1}

\caption{
Spatial distributions of the field enhancement factors (a, b) and phases (c, d)
of the horizontal and the perpendicular field.
}\label{fig_2}

\caption{
The measured (a) and the simulated (b) terahertz response as a function of $V_{DS}$ and $V_G$.
Three vertical lines are marked at $V_S=2$~V, $V_D=V_S=0$~V, and $V_D=2$~V.
Along the line at $V_D=2$, five bias conditions are marked by
$\alpha$, $\beta$, $\gamma$, $\delta$, and $\chi$, corresponding to
$V_G=-1.0$, $-2.0$, $-3.0$, $-3.5$, and $-4.1$~V, respectively.
In the SR regime when the bias is applied at the source,
there is a clear deviation between the experiment and the simulation.
}\label{fig_3}

\caption{
(a) The spatial distributions of the charge modulation $\mathrm{d}n/\mathrm{d}V_{GE}$
and the self-mixing term $\dot{\xi_x}\dot{\xi_z}\cos\phi$ at $V_G=-4.1$~V
with a DC bias of 2~V applied at the source or at the drain.
(b) The spatial distributions of charge modulation at different gate voltages
and at a constant drain bias ($V_D=2.0$~V).
The mixing current as a function of $V_G$ with
(c) $V_D=V_S=0$~V, (d) $V_D=2.0$~V, and (e) $V_S=2.0$~V.
}\label{fig_4}

\end{figure}

\newpage
\begin{figure}[!p]
\includegraphics[width=.8\textwidth]{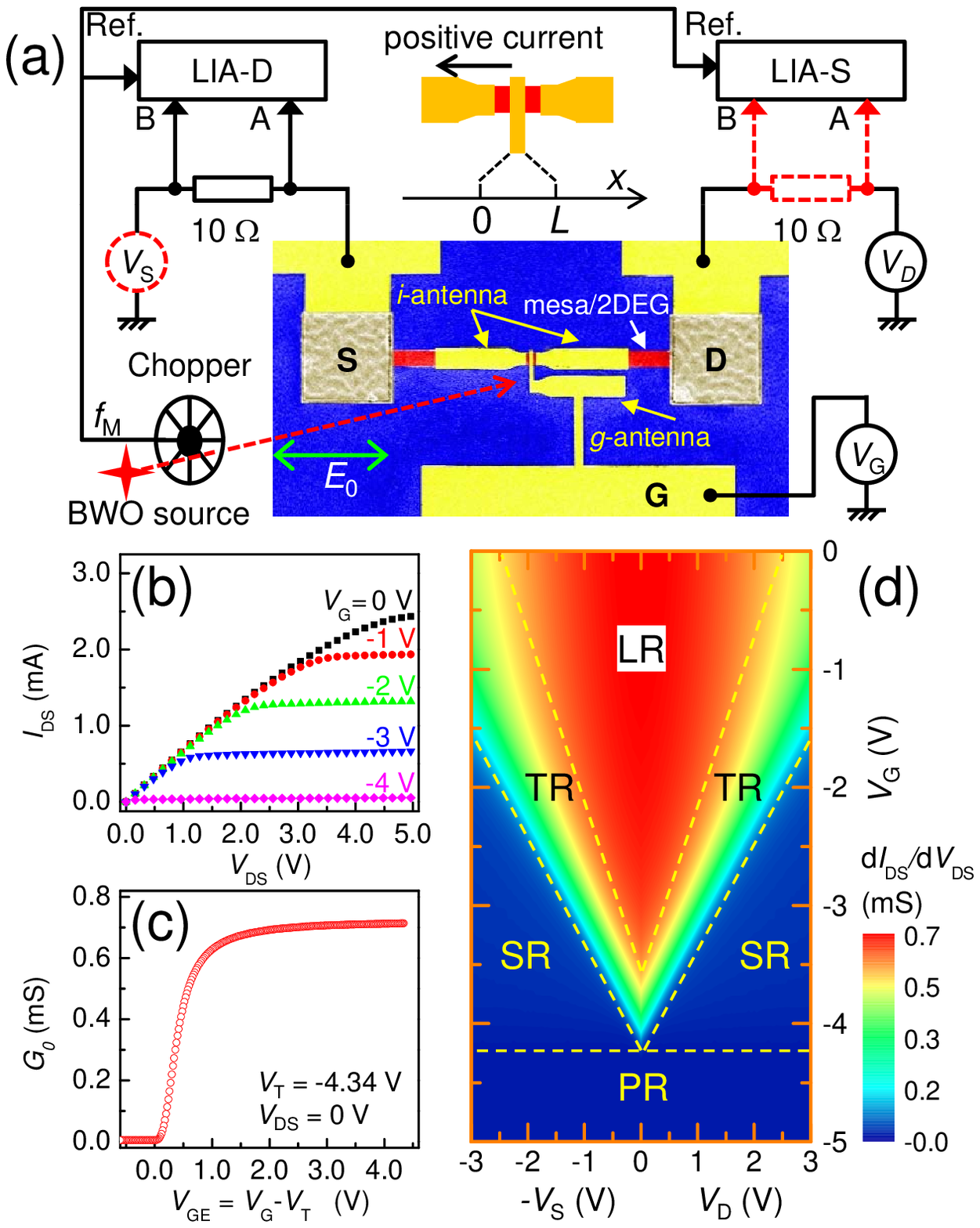}
\center{Sun {\it et al:~} Figure 1/4}
\end{figure}
\begin{figure}[!p]
\includegraphics[width=.8\textwidth]{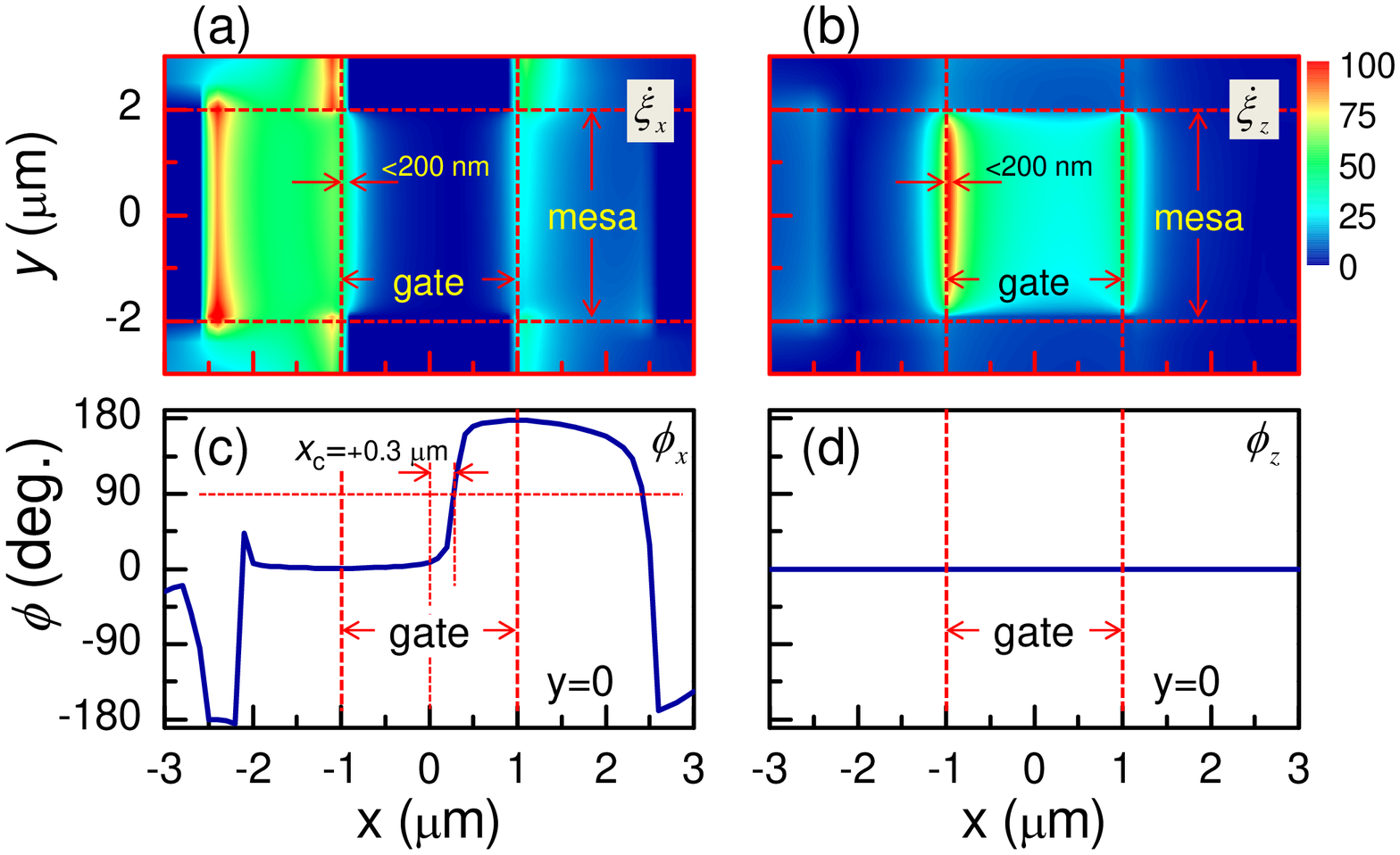}
\center{Sun {\it et al:~} Figure 2/4}
\end{figure}
\begin{figure}[!htp]
\includegraphics[width=.8\textwidth]{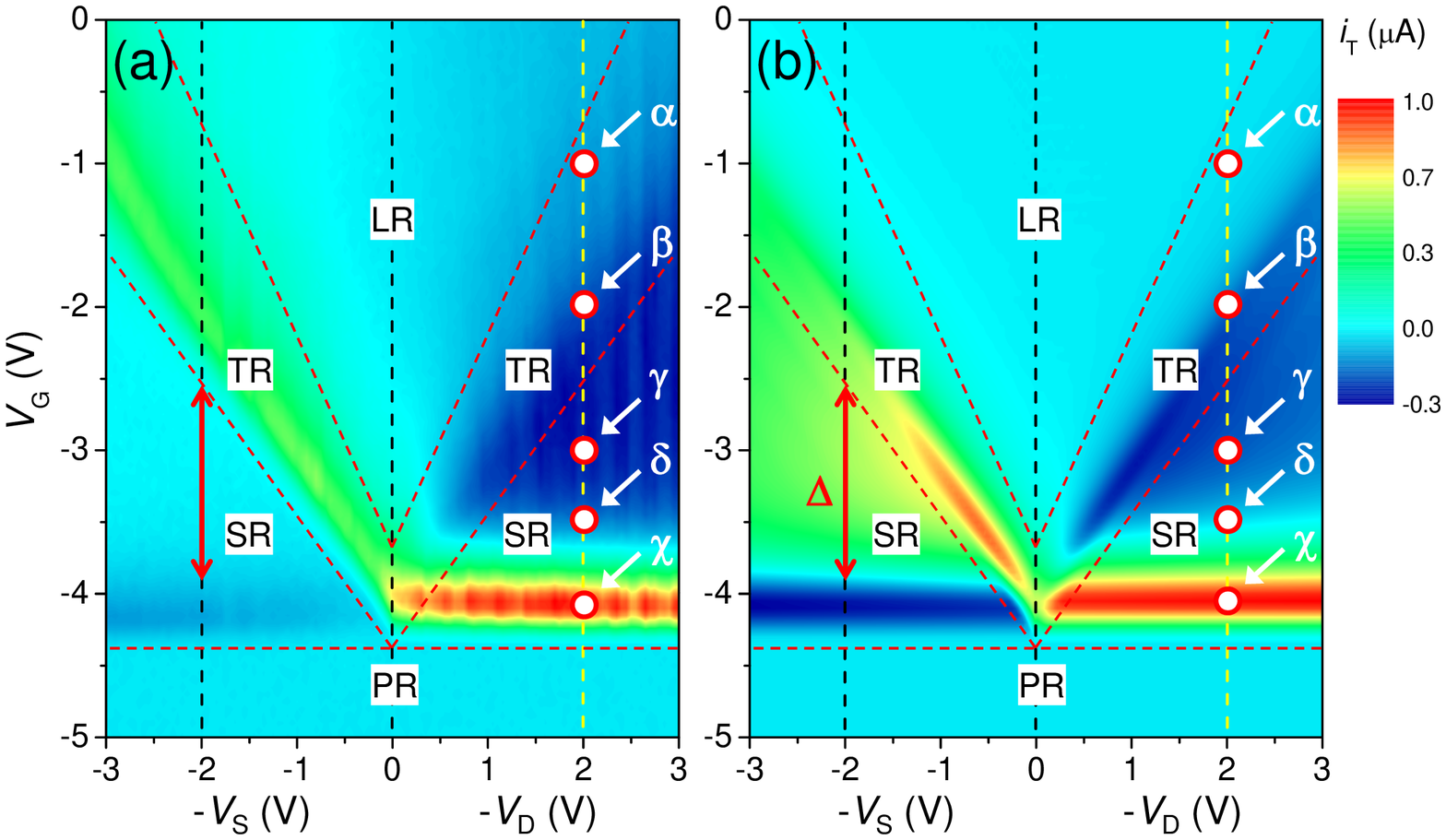}
\center{Sun {\it et al:~} Figure 3/4}
\end{figure}
\begin{figure}[!p]
\includegraphics[width=.8\textwidth]{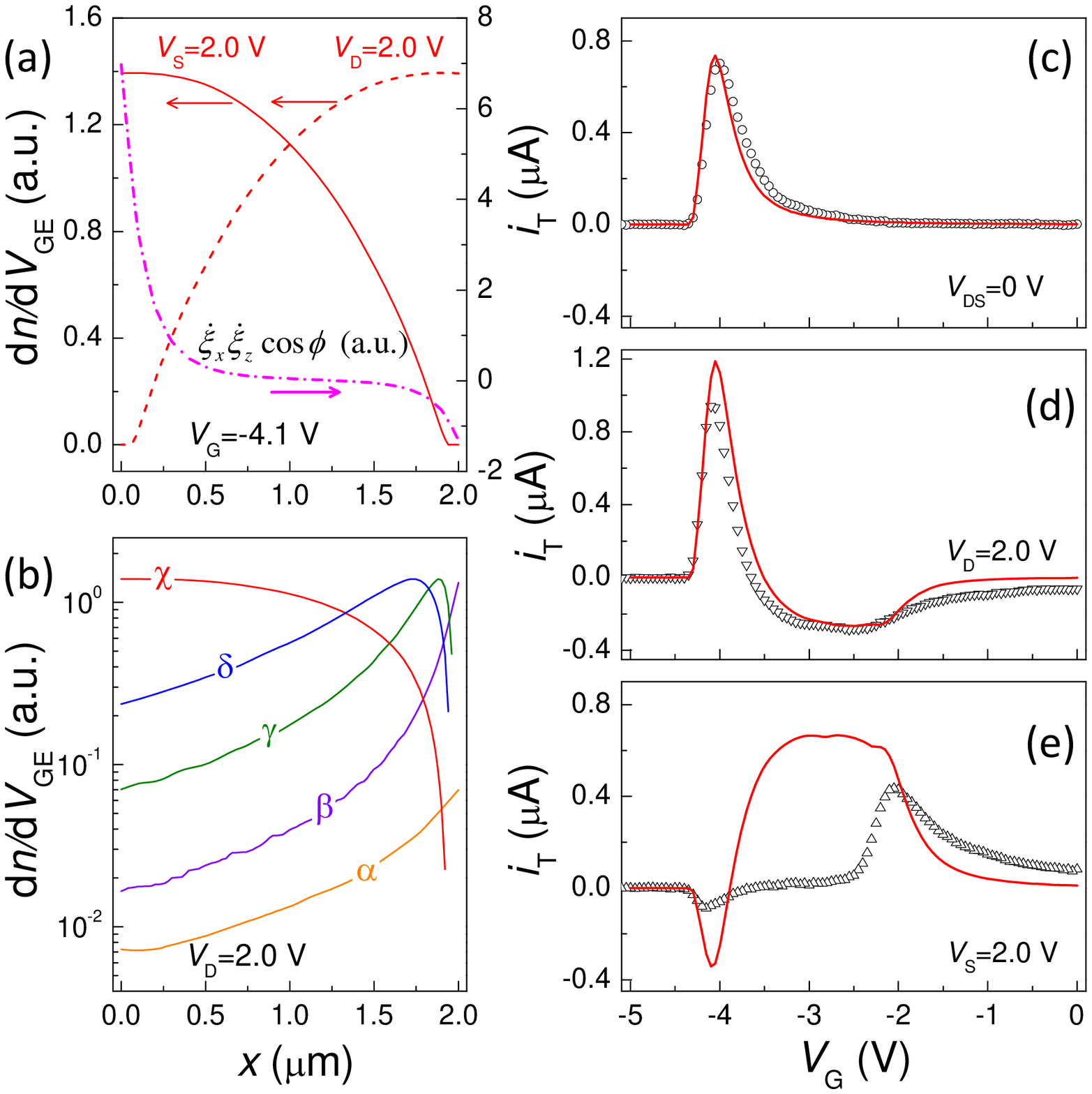}
\center{Sun {\it et al:~} Figure 4/4}
\end{figure}


\begin{thebibliography}{99}
\bibitem{note_1}
Similar terahertz responses have been observed (but not published) in other types of field-effect detectors,
such as the nanometer-sized FETs and a quantum-point-contact device with symmetric bow-tie antenna.
\bibitem{note_2}
In the term $i_{xx}$, the terahertz potential $\xi_xE_0$ is
obtained by integrating the horizontal field from $x_c$ to $x=0$ and from $x_c$ to $x=L$ by setting $\xi_x(x_c)=0$.
The ratio of the horizontal field to the perpendicular field ranges
from 0.6 at $x=0$ and 0.5 at $x=L$ to 0 at $x=x_c$.
The mean value of the ratio is about 0.19. Hence, the self-mixing from term $i_{xx}$ is a minor contribution.
\end{thebibliography}
\end{document}